\documentclass[notitlepage,aps,floats,showpacs,showkeys,preprintnumbers,nofootinbib]{revtex4-1}

\pdfoutput=1

\usepackage{amsmath}
\usepackage{amssymb}
\usepackage{epsfig}
\usepackage{bbm}

\usepackage{epstopdf}

\def\mueff{\mu_{\text{eff}}}
\def\fbi{~{\rm fb}^{-1}}

\def\pb{~{\rm pb}}

\def\ie{{\it i.e.}}
\def\anti{\overline}
\def\sigsi{\sigma_{\rm SI}}

\def\gev{~{\rm GeV}}
\def\tev{~{\rm TeV}}
\def\beq{\begin{equation}}
\def\eeq{\end{\equation}}
\def\bea{\begin{eqnarray}}
\def\eea{\end{eqnarray}}
\def\gam{\gamma}
\def\kap{\kappa}
\def\lam{\lambda}
\def\akap{A_\kap}
\def\alam{A_\lam}
\def\mhalf{m_{1/2}}

\def\mhusq{m_{H_u}^2}
\def\mhdsq{m_{H_d}^2}
\def\mssq{m_{S}^2}
\def\omghsq{\Omega h^2}
\def\amu{a_\mu}
\def\damu{\delta \amu}
\def\br{{\rm BR}}

\def\hsm{h_{\rm SM}}

\def\to{\rightarrow}
\def\hi{h_1}
\def\hii{h_2}
\def\hiii{h_3}
\def\hpm{H^\pm}
\def\ai{a_1}
\def\aii{a_2}
\def\mhi{m_{\hi}}
\def\mhii{m_{\hii}}
\def\mai{m_{\ai}}
\def\maii{m_{\aii}}
\def\mhiii{m_{\hiii}}
\def\mhpm{m_{\hpm}}
\def\mh{m_h}

\def\cnone{\widetilde \chi_1^0}

\def\mcnone{m_{\cnone}}

\def\hsm{h_{\rm SM}}

\def\msq{m_{\tilde q}}
\def\mgl{m_{\tilde g}}
\def\stopi{\tilde t_1}
\def\stopii{\tilde t_2}
\def\mstopii{m_{\stopii}}
\def\mstopi{m_{\stopi}}

\def\tanb{\tan\beta}
\def\cotb{\cot\beta}
\def\beq{\begin{equation}}
\def\eeq{\end{equation}}

\def\rhigg#1{R_{gg}^{\hi}(#1)}
\def\rhiigg#1{R_{gg}^{\hii}(#1)}

\def\sig{\sigma}
\def\sigres{\sig_{\text{res}}}

\def\lsim{\mathrel{\raise.3ex\hbox{$<$\kern-.75em\lower1ex\hbox{$\sim$}}}}
\def\gsim{\mathrel{\raise.3ex\hbox{$>$\kern-.75em\lower1ex\hbox{$\sim$}}}}

\def\bit{\begin{itemize}}
\def\eit{\end{itemize}}
\def\bec{\begin{center}}
\def\eec{\end{center}}

\begin{document}
\title{Could two NMSSM Higgs bosons be present near 125 GeV?}

\author{John F.~Gunion}\email{jfgunion@ucdavis.edu}
\author{Yun~Jiang}\email{yunjiang@ucdavis.edu}
\affiliation{\,Department of Physics, University of California, Davis, CA 95616, USA}
\author{Sabine~Kraml}\email{sabine.kraml@lpsc.in2p3.fr}
\affiliation{\,Laboratoire de Physique Subatomique et de Cosmologie, UJF Grenoble 1, CNRS/IN2P3, INPG, 
53 Avenue des Martyrs, F-38026 Grenoble, France}


\begin{abstract}
We examine GUT-scale NMSSM  scenarios in which {\it both} $\hi$ and $\hii$ lie in the $123\mbox{--}128\gev$ mass range.  Very substantially enhanced $\gam\gam$ and other rates are possible.  Broadened mass peaks are natural. 
\end{abstract}


\maketitle


Data from the ATLAS and CMS collaborations~\cite{HiggsDiscovery,atlashiggs,cmshiggs}
provide an essentially $5\sigma$ signal for a Higgs-like resonance with mass of order $123\mbox{--}128\gev$. 
In the $\gam\gam$ final state, the ATLAS and CMS rates are roughly $1.9\pm0.4$ and $1.6\pm 0.4$  times the Standard Model (SM) prediction. In the $ZZ\to 4\ell$ channel, the ATLAS and CMS signals are roughly $1.1^{+0.5}_{-0.4}$ and $0.7^{+0.4}_{-0.3}$ times the SM expectation, respectively.
In the $b\anti b$, $\tau^+\tau^-$ and $WW\to \ell\nu\ell\nu$ channels, the central value ATLAS rates are somewhat suppressed relative to the SM prediction but error bars are very large. The CMS signals in these latter channels are also somewhat suppressed and lie at least $1\sigma$ below the SM prediction --- no signal being observed in the $\tau^+\tau^-$ channel. 
 Meanwhile, the CDF and D0 experiments have announced new results \cite{newtevatron} that support the $\sim 125\gev$ Higgs signal and suggest an enhancement relative to the SM of the $W$+Higgs with Higgs$\to b\anti b$ rate by a factor of $2\pm 0.6$.

Enhanced rates in the $\gam\gam$ channel  have been shown to be difficult to achieve in the NMSSM \cite{Gunion:2012zd}, while remaining consistent with all relevant constraints, including those from LEP  searches, $B$-physics, the muon anomalous magnetic moment,
$\amu\equiv (g-2)_\mu/2$, and the relic density of dark matter, $\omghsq$,
when parameters are semi-unified at the GUT scale. By ``semi-unified'' we mean a model in which $m_0$, $\mhalf$, and $A_0$ are universal at the GUT scale with NUHM relaxation for $\mhusq$, $\mhdsq$ and $\mssq$ and general $\alam$ and $\akap$. Enhancements appear to be possible only if large values of the superpotential coupling $\lam$ are employed and the $\amu$ constraint is greatly relaxed \cite{Ellwanger:2012ke}. (See \cite{Ellwanger:2011aa} for the first discussion of an enhanced $\gam\gam$ rate at large $\lam$ in the NMSSM with parameters defined at the weak scale.)
In this Letter,  we pursue the case of generally large $\lam$ and uncover a particularly interesting set of scenarios in which the two lightest CP-even Higgs bosons, $\hi$ and $\hii$, both lie in the $123\mbox{--}128\gev$ mass window.  Phenomenological consequences are examined.


For the numerical analysis, we use NMSSMTools~\cite{Ellwanger:2004xm}\cite{Ellwanger:2005dv}\cite{nmweb}  version 3.2.0, which has improved convergence of RGEs in the case of large Yukawa couplings and thus allows us to explore parameter regions that where left uncharted in~\cite{Gunion:2012zd}. 
The precise constraints imposed are the following.  Our `basic constraints' will be to require
that an NMSSM parameter choice be such as to give a proper RGE solution, have no Landau pole, have a neutralino as the lightest SUSY particle (LSP) and obey Higgs and SUSY mass limits as implemented 
in NMSSMTools-3.2.0 (Higgs mass limits are from LEP, older TEVATRON, and early LHC data; SUSY mass limits are essentially from LEP.) 

Regarding $B$ physics, the constraints considered are those on $\br(B_s\to X_s\gamma)$, $\Delta M_s$, $\Delta M_d$, $\br(B_s\to \mu^+\mu^-)$, $\br(B^+\to \tau^+\nu_\tau)$ 
and $\br(B\to X_s \mu^+ \mu^-)$ at $2\sigma$ as encoded in
NMSSMTools-3.2.0, except that we updated the 
bounds on rare $B$ decays
to $3.04<\br(B_s\to X_s\gamma)\times 10^{4}<4.06$ and 
$\br(B\to\mu^+\mu^-)<4.5\times 10^{-9}$; 
theoretical uncertainties in $B$-physics observables are taken into
account as implemented in NMSSMTools-3.2.0.  

Regarding dark matter constraints, we accept all points that have $\omghsq<0.136$, 
thus allowing for scenarios in which the relic density arises at least in part from some other source. 
However, we single out points with $0.094\leq\omghsq\leq0.136$, which is the
`WMAP window' defined in NMSSMTools-3.2.0 after including 
theoretical and experimental systematic uncertainties.  
In addition, we impose bounds on the spin-independent LSP--proton scattering cross section implied by  the neutralino-mass-dependent Xenon100 bound~\cite{Aprile:2011hi}. (For points with $\omghsq<0.094$, we rescale these bounds by a factor of $0.11/\omghsq$.)

Our  study focuses in particular on NMSSM parameter choices such that both $\mhi$ and $\mhii$ lie within $123\mbox{--}128\gev$. 
We focus moreover on $\lam\geq 0.1$, a range for which it is known~\cite{Ellwanger:2012ke}\cite{Ellwanger:2011aa} that some enhancement, relative to the SM, of the Higgs signal in the $\gam\gam$ final state is possible.  The degenerate situation is especially interesting in that 
an enhanced $\gam\gam$ rate at $\sim 125\gev$ could arise as a result of the $\hi$ and $\hii$ rates summing together, even if the individual rates are not full SM-like strength (or enhanced).

Above, we did not mention imposing a constraint on $\amu$. Rough consistency with the measured value of $\amu$  requires that the extra NMSSM contribution, $\delta \amu$, falls into 
the window defined in NMSSMTools of $8.77\times 10^{-10}<\delta\amu<
4.61\times 10^{-9}$ expanded to $5.77\times
10^{-10}<\delta\amu<4.91\times 10^{-9}$ after allowing for a $1\sigma$
theoretical error in the NMSSM calculation of $\pm 3\times
10^{-10}$. In fact, given the previously defined constraints and focusing on $\lam\geq 0.1$, $\damu$ is always 
too small, being at most $\sim 2\times 10^{-10}$.  Demanding $\damu$ large enough to fall into the above window, or even come close to doing so, appears from our scans to date to only be possible if $\lam<0.1$ \cite{Gunion:2012zd}, for which the Higgs signal in the $\gam\gam$ and $VV^*$ ($V=W,Z$) final states for Higgs in the $123\mbox{--}128\gev$ window is very SM-like.

The main production/decay channels relevant for current LHC data are gluon-gluon and $WW$ fusion to Higgs with Higgs decay to $\gam\gam$ or $ZZ^*\to 4\ell$. The LHC is also beginning to probe $W,Z+$Higgs with Higgs decay to $b\anti b$, a channel for which Tevatron data is relevant, and $WW\to$Higgs with Higgs$\to\tau^+\tau^-$. 
For the cases studied, where there are two nearly degenerate Higgs bosons, we will combine their signals as follows in defining the mass and signal for the effective Higgs, $h$.  First, for the individual Higgs we compute the ratio of the $gg$ or $WW$-fusion (VBF) induced Higgs cross section times the Higgs branching ratio to a given final state, $X$, relative to the corresponding value for the SM Higgs boson:
\begin{eqnarray}
R^{h_i}_{gg}(X)&\equiv& {\Gamma(gg\to h_i) \ \br(h_i\to X)\over \Gamma(gg\to \hsm)\ \br(\hsm\to X)}, \\[2mm]
R^{h_i}_{\rm VBF}(X)&\equiv& {\Gamma(WW\to h_i) \ \br(h_i\to X)\over \Gamma(WW\to \hsm)\ \br(\hsm\to X)},
\end{eqnarray} 
where $h_i$ is the $i^{th}$ NMSSM scalar Higgs, and $\hsm$ is the SM
Higgs boson. Note that the corresponding ratio for  $V^*\to V h_i$ ($V=W,Z$) with $h_i\to X$ is equal to $R^{h_i}_{\rm VBF}(X)$. These ratios are computed in a self-consistent manner (that is,
treating radiative corrections for the SM Higgs boson in the same
manner as for the NMSSM Higgs bosons) using an
appropriate additional routine for the SM Higgs added to the NMHDECAY
component of the NMSSMTools package.   Next, we compute the effective Higgs mass in given production and final decay channels $Y$ and $X$, respectively, as
\beq
m_h^Y(X)\equiv {R^{\hi}_{Y}(X) \mhi +R^{\hii}_Y(X) \mhii \over R^{\hi}_{Y}(X)  +R^{\hii}_Y(X) }\,
\eeq
and define the net signal to simply be 
\beq
R^h_Y(X)=R^{\hi}_Y(X)+ R^{\hii}_Y(X)\,.
\eeq
Of course, the extent to which it is appropriate to combine the rates from the $\hi$ and $\hii$ depends upon the degree of degeneracy and the experimental resolution.  For the latter, we assume 
$\sigres \sim 1.5\gev$~\cite{Chatrchyan:2012tw}.\footnote{The values for $\sigres$ quoted in this paper range from  $1.39$--$1.84\gev$ to $2.76$--$3.19\gev$, the better resolutions being for the case where both photons are in the barrel and the worse resolutions for when one or both photons are in the endcap.  We anticipate that the more recent analyses have achieved substantially better mass resolutions, but details are not yet available.} 
It should be noted that the widths of the $\hi$ and $\hii$ are of the same order of magnitude as the width of a 125 GeV SM Higgs boson, \ie\ they  are very much smaller than this resolution. 



We perform scans covering the following parameter ranges, which correspond to an expanded version of those considered in \cite{Ellwanger:2012ke}: 
$0 \le m_0 \le 3000$; 
$100 \le m_{1/2} \le 3000$;
$1 \le \tan\beta \le 40 $; 
$-6000 \le A_0 \le 6000 $; 
$0.1 \le \lambda \le 0.7$;  
$0.05 \le \kappa \le 0.5$; 
$-1000 \le A_\lambda \le 1000$; 
$-1000 \le A_\kappa \le 1000$; 
$100 \le \mu_{\text{eff}} \le 500$.
In the figures shown in the following, we only display points which pass the basic constraints, satisfy $B$-physics constraints, have $\Omega h^2<0.136$, obey the XENON100 limit on the LSP scattering cross-section off protons {\it and} have {\it both} $\hi$ and $\hii$  in the desired mass range: $123\gev <\mhi,\mhii<128\gev$. 

\begin{figure}[h!]
\begin{center}
\hspace*{-.5in}\includegraphics[width=0.75\textwidth]{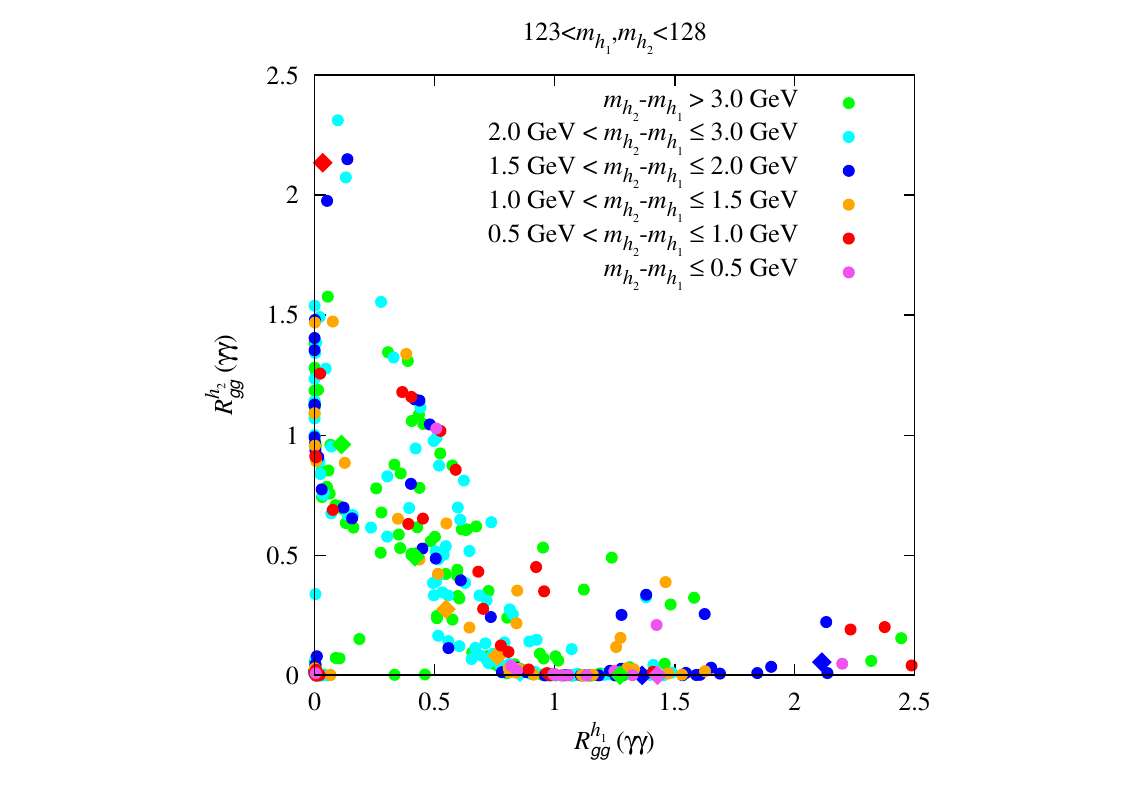}
\end{center}\vspace*{-.2in}
\caption{Correlation of $gg\to (\hi,\hii)\to \gam\gam$ signal strengths when both $\hi$ and $\hii$ lie in the $123\mbox{--}128\gev$ mass range. The circular points have $\omghsq<0.094$, while diamond points have $0.094\leq \omghsq\leq 0.136$.  
Points are color coded according to $\mhii-\mhi$ as indicated on the figure. }
\label{rgg12}
\end{figure}

In Fig.~\ref{rgg12}, we display $R_{gg}^{\hii}(\gam\gam)$ versus $R_{gg}^{\hi}(\gam\gam)$ with points color coded according to $\mhii-\mhi$. The circular points have $\omghsq<0.094$, while diamond points have $0.094\leq \omghsq\leq 0.136$ (within the WMAP window). We observe a large number of points for which $\mhi,\mhii\in[123,128]\gev$ and many are such that $\rhigg{\gam\gam}+\rhiigg{\gam\gam}>1$. A few such points have $\omghsq$ in the WMAP window. These points are such that either $\rhigg{\gam\gam}>2$ or $\rhiigg{\gam\gam}>2$, with the $R$ for the other Higgs being small. However, the majority of the points with $\rhigg{\gam\gam}+\rhiigg{\gam\gam}>1$ have $\omghsq$ below the WMAP window and for many the $\gam\gam$ signal is shared between the $\hi$ and the $\hii$.

Based on these results, we will now combine the $\hi$ and $\hii$ signals as described above and present plots coded according to the following legend. First, we note that circular (diamond) points have $\omghsq<0.094$ ($0.094\leq \omghsq\leq 0.136$). We then color the points according to:
\begin{itemize}
\item 
red for $\mhii-\mhi\leq 1\gev$;
\item  
blue for $1\gev <\mhii-\mhi\leq 2\gev$;
\item 
green for  $2\gev<\mhii-\mhi\leq 3\gev$.
\end{itemize}
For current statistics and $\sig_{\rm res}\gsim 1.5\gev$ we estimate that the $\hi$ and $\hii$ signals will not be seen separately for $\mhii-\mhi\leq 2\gev$.

\begin{figure}[ht]
\begin{center}
\hspace*{-.2in}\includegraphics[width=0.54\textwidth] {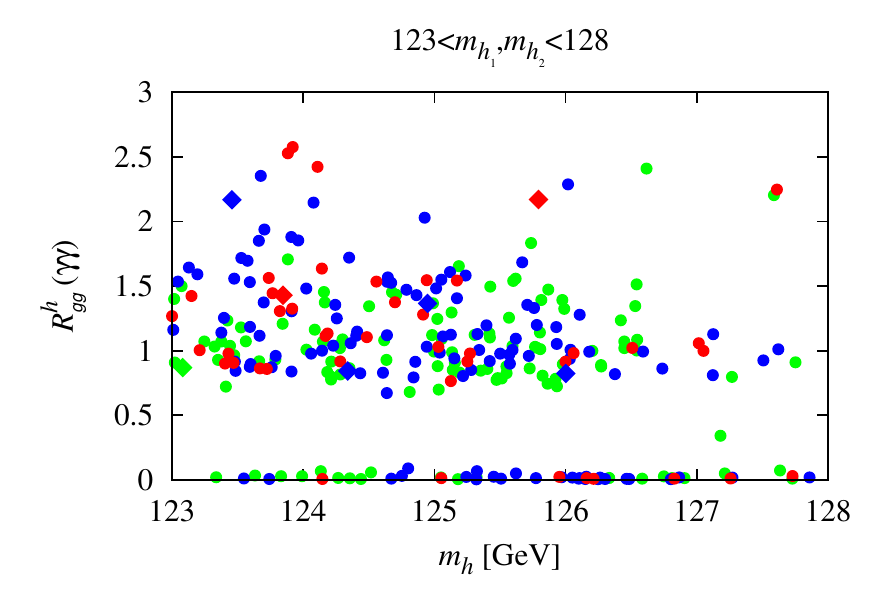}\hspace*{-.2in}\includegraphics[width=0.54\textwidth] {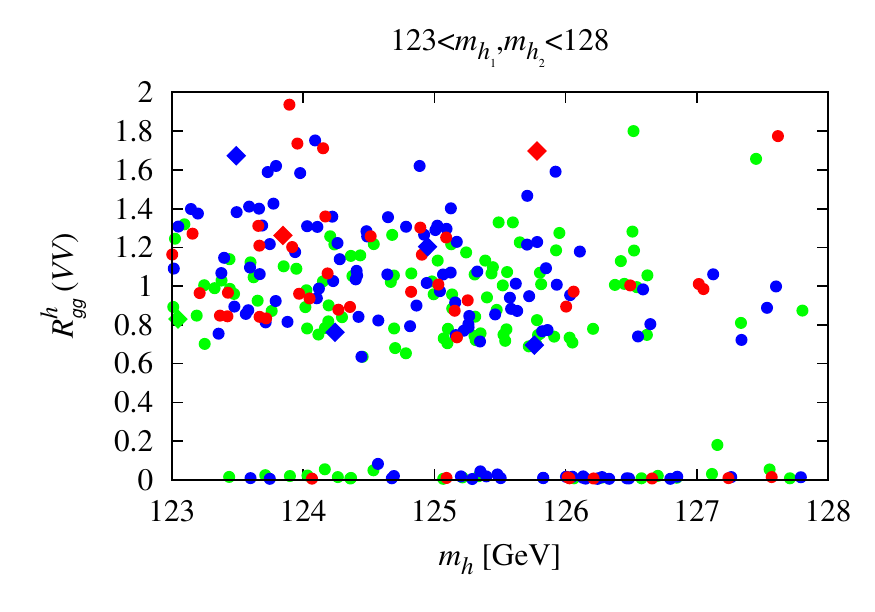}
\hspace*{-.2in}\includegraphics[width=0.54\textwidth] {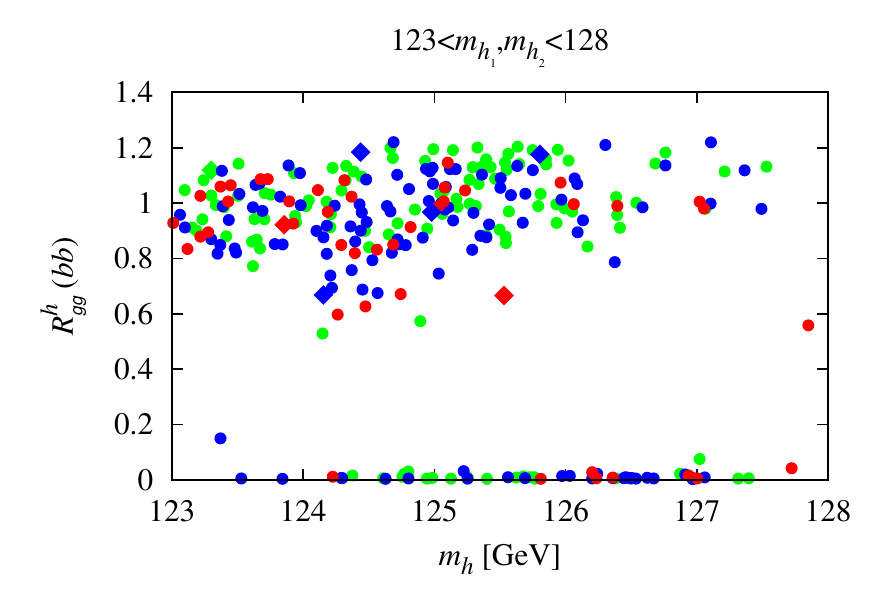}\hspace*{-.2in}\includegraphics[width=0.54\textwidth] 
{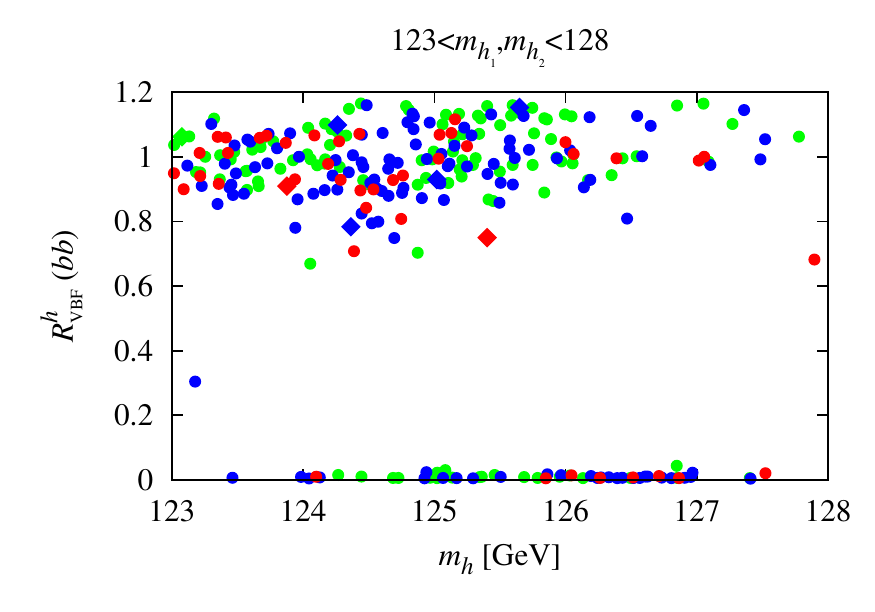}
\end{center}\vspace*{-.15in}
\caption{$R^h_{gg}(X)$  for $X=\gam\gam,VV,b\anti b$, and $R^h_{\rm VBF}(b\anti b)$  versus $m_h$. For application to the Tevatron, note that $R_{\rm VBF}^h(b\anti b)=R^h_{W^*\to Wh}(b\anti b)$.
The color code here and in the following figures is green for points with $2\gev<\mhii-\mhi\leq 3\gev$, blue for $1\gev <\mhii-\mhi\leq 2\gev$, and red for $\mhii-\mhi\leq 1\gev$. }
\label{rgg2}
\end{figure}

In Fig.~\ref{rgg2} we show results for $R_{gg}^h(X)$ with $\mh\in[123,128]\gev$ as a function of $\mh$ for $X=\gam\gam,VV,b\bar b$.  Enhanced $\gam\gam$ and $VV$ rates from gluon fusion are very common. The bottom-right plot shows that enhancement in the $Wh$ with $h\to b\anti b$ rate is also natural, though not as large as the best fit value suggested by the new Tevatron analysis \cite{newtevatron}. 
Diamond points (\ie\ those in the WMAP window) are rare, but typically show enhanced rates. 


\begin{figure}[ht]
\begin{center}
\hspace*{-.23in}\includegraphics[width=0.53\textwidth]{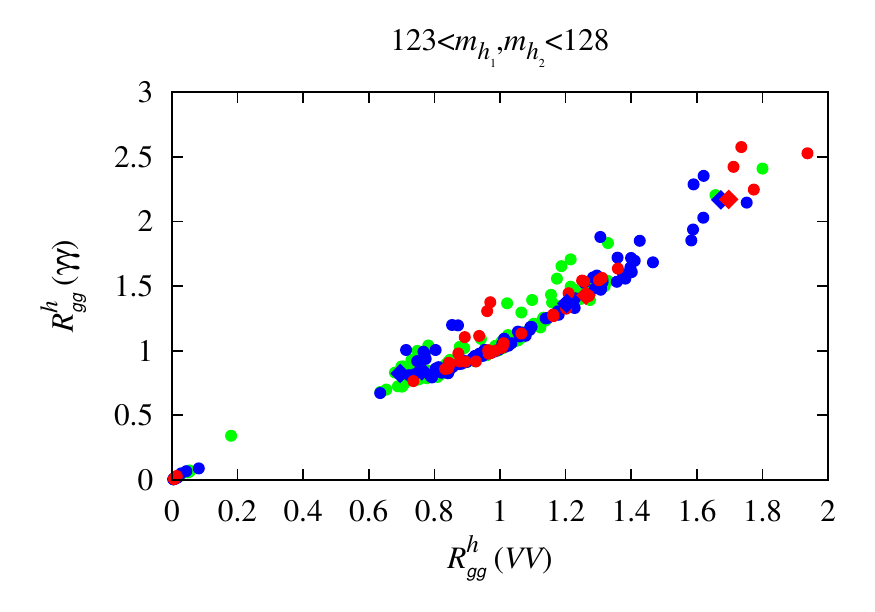}\hspace*{-.22in}\includegraphics[width=0.53\textwidth]{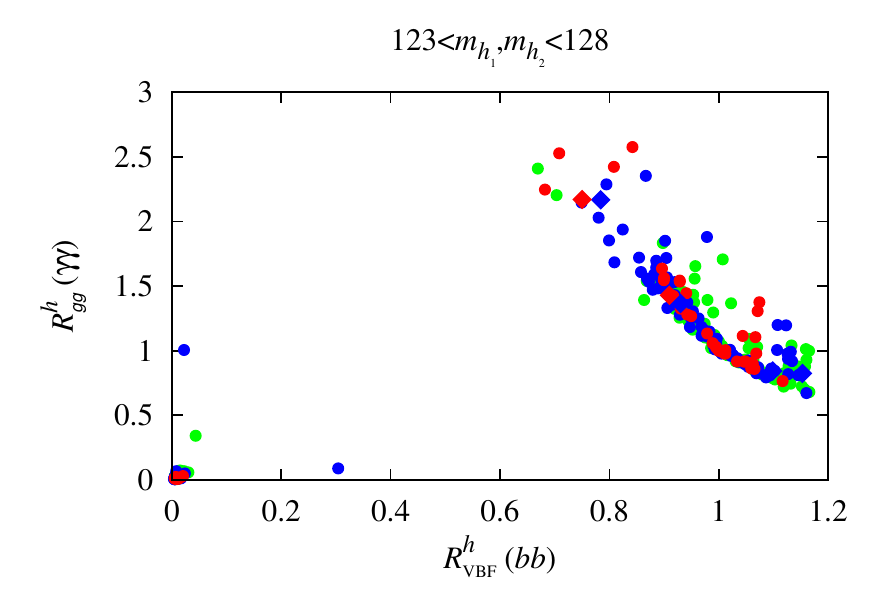}
\end{center}\vspace*{-.1in}
\caption{Left: correlation between the gluon fusion induced $\gam\gam$ and $VV$ rates relative to the SM.  
Right: correlation between the gluon fusion induced $\gam\gam$ rate and the $WW$ fusion induced $b\anti b$ rates relative to the SM; the relative rate for $W^*\to Wh$ with $h\to b\anti b$ (relevant for the Tevatron) is equal to the latter.}
\label{correl2}
\end{figure}

In Fig.~\ref{correl2}, we display in the left-hand plot the strong correlation between $R^h_{gg}(\gam\gam)$ and $R^h_{gg}(VV)$.  Note that if $R^h_{gg}(\gam\gam)\sim 1.5$, as suggested by current experimental results, then in this model  $R^h_{gg}(VV)\geq 1.2 $. The right-hand plot shows the (anti) correlation between $R^h_{gg}(\gam\gam)$  and $R_{W^*\to Wh}^h(b\anti b)=R_{\rm VBF}^h(b\anti b)$.  In general, the larger $R^h_{gg}(\gam\gam)$ is, the smaller the value of $R_{W^*\to Wh}^h(b\anti b)$. However, this latter plot shows that there {\it are} parameter choices for which both the $\gam\gam$ rate at the LHC and the $W^*\to Wh(\to b\anti b)$ rate at the Tevatron (and LHC) can be enhanced relative to the SM as a result of there being contributions to these rates from both the $\hi$ and $\hii$. It is often the case that one of the $\hi$ or $\hii$ dominates $R^h_{gg}(\gam\gam)$ while the other dominates $R_{W^*\to Wh}^h(b\anti b)$.  This is typical of the diamond WMAP-window points.  However, a significant number of the circular $\omghsq<0.094$ points are such that either the $\gam\gam$ or the $b\anti b$ signal receives substantial contributions from both the $\hi$ and the $\hii$ (as seen, for example,  in Fig.~\ref{rgg12}  for the $\gam\gam$ final state) while the other final state is dominated by just one of the two Higgses.  We did not find points where the $\gam\gam$ and $b\anti b$ final states {\it both} receive substantial contributions from {\it both} the $\hi$ and $\hii$.

\begin{figure}[ht]
\begin{center}
\hspace*{-.23in}\includegraphics[width=0.53\textwidth]{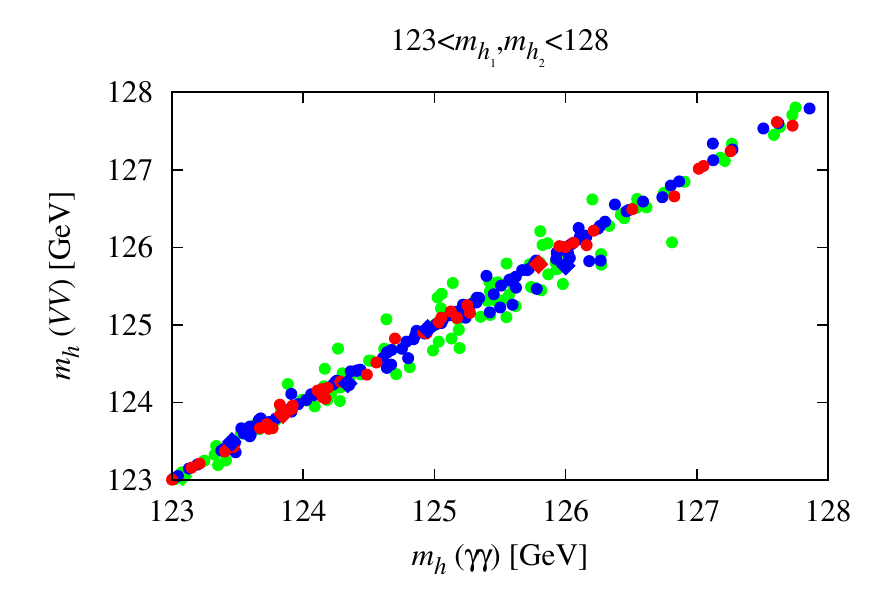}\hspace*{-.22in}\includegraphics[width=0.53\textwidth]{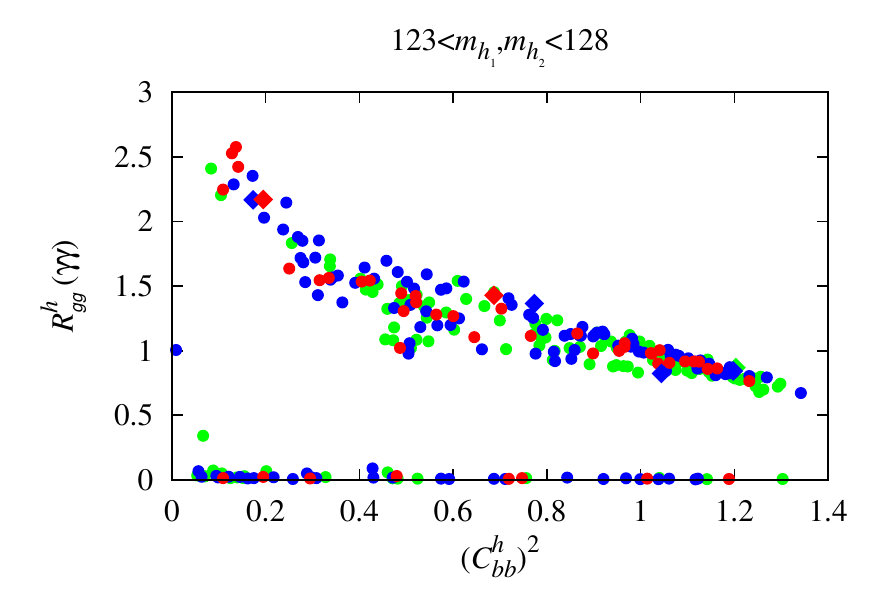}
\end{center}\vspace*{-.1in}
\caption{Left: effective Higgs masses obtained from different channels: $m_h^{gg}(\gam\gam)$ versus  $m_h^{gg}(VV)$.  
Right: $\gam\gam$ signal strength $R_{gg}^{h}(\gam\gam)$ versus effective coupling to $b\bar b$ quarks $({C^h_{b\bar b}})^2$.  Here, ${C^h_{b\bar b}}^2\equiv\left[R_{gg}^{\hi}(\gam\gam){C^{\hi}_{b\bar b}}^2 +R_{gg}^{\hii} (\gam\gam){C^{\hii}_{b\bar b}}^2\right]/ \left[R_{gg}^{\hi}(\gam\gam)+R_{gg}^{\hii} (\gam\gam)\right]$.}
\label{rgginfo}
\end{figure}

As noted above, there is a strong correlation between $R^h_{gg}(\gam\gam)$ and $R^h_{gg}(VV)$ described approximately by $R^h_{gg}(\gam\gam) \sim 1.25\, R^h_{gg}(VV)$.
Thus, it is not surprising that  the  $m_h$ values for the gluon fusion induced $\gam\gam$ and $VV$ cases are also strongly correlated --- in fact, they differ by no more than a fraction of a GeV and are most often much closer, see the left plot of Fig.~\ref{rgginfo}.  The right plot of Fig.~\ref{rgginfo} illustrates the mechanism behind enhanced rates, namely that large net $\gam\gam$ branching ratio is achieved by reducing the average total width by reducing the average $b\anti b$ coupling strength.

\begin{figure}[ht]
\begin{center}
\hspace*{-.23in}\includegraphics[width=0.52\textwidth]{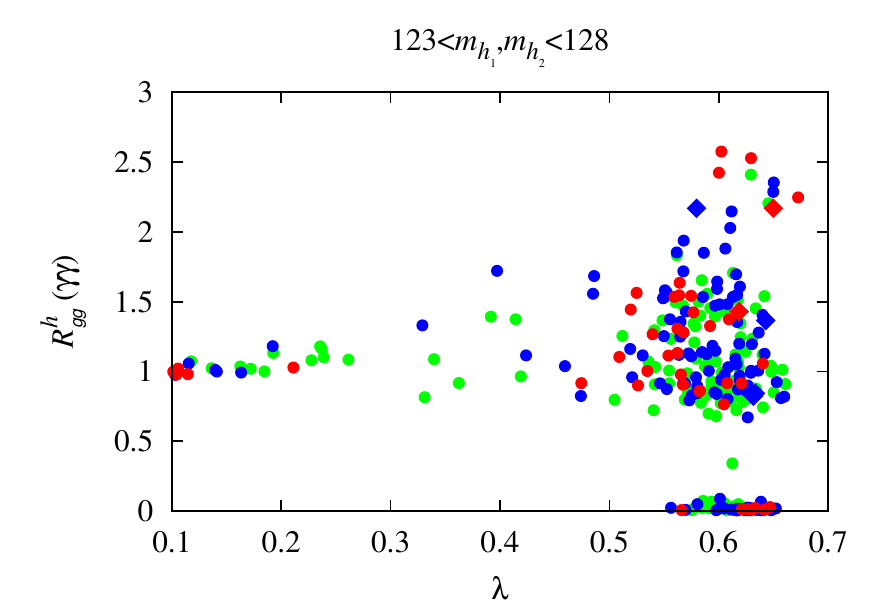}\hspace*{-.2in}\includegraphics[width=0.52\textwidth]{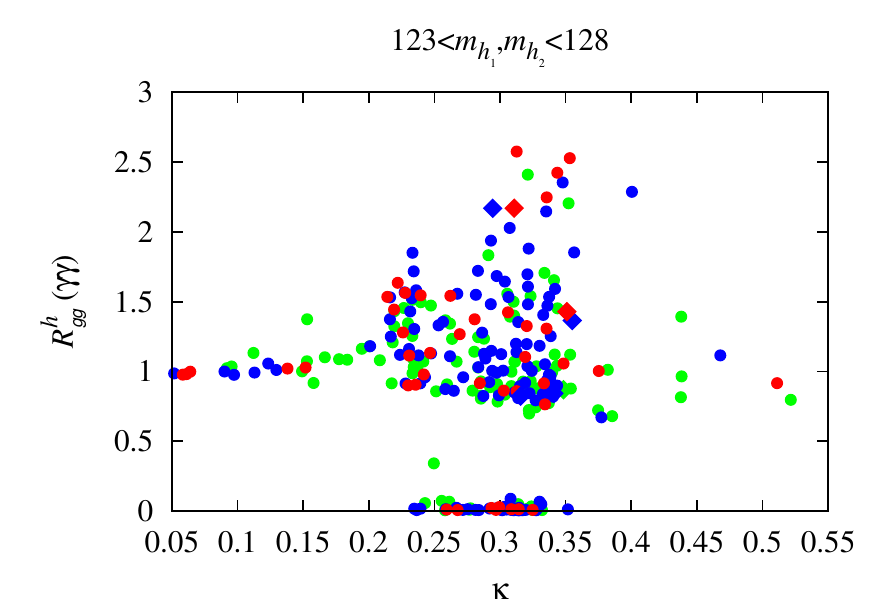}
\hspace*{-.23in}\includegraphics[width=0.52\textwidth]{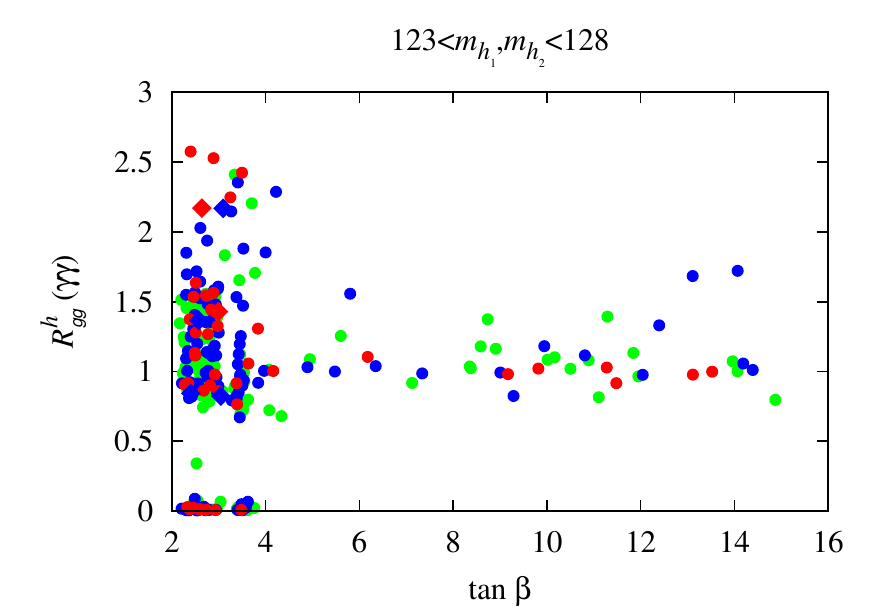}\hspace*{-.2in}\includegraphics[width=0.52\textwidth]{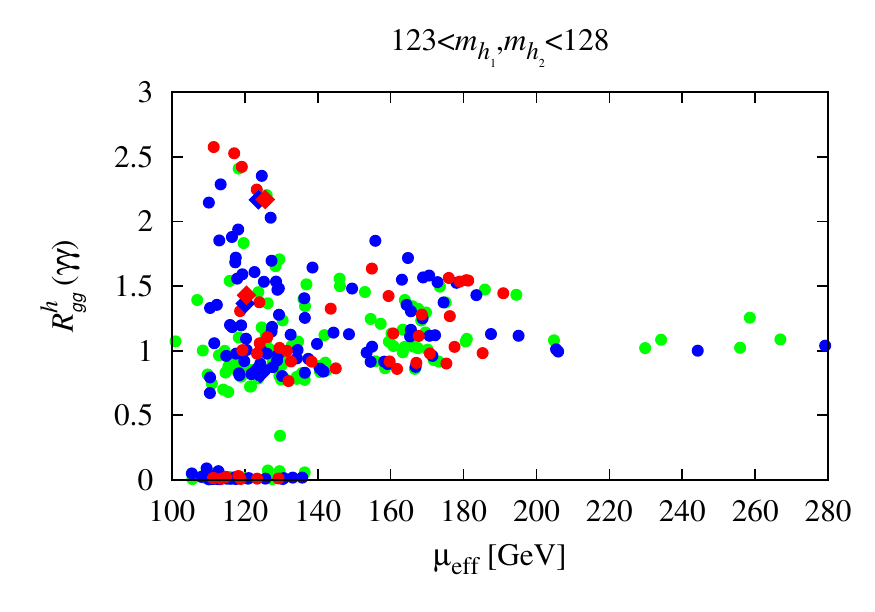}
\end{center}\vspace*{-.1in}
\caption{Dependence of $R^h_{gg}(\gam\gam)$ on $\lam$, $\kap$, $\tanb$ and $\mueff$.}
\label{rvsall}
\end{figure}

The dependence of $R^h_{gg}(\gam\gam)$ on $\lam$, $\kap$, $\tanb$ and $\mueff$ is illustrated in Fig.~\ref{rvsall}.  We observe that the largest $R^h_{gg}(\gam\gam)$ values arise at large $\lam$, moderate $\kap$, small $\tanb<5$ (but note that $R^h_{gg}(\gam\gam)>1.5$ is possible even for $\tanb=15$) and small $\mueff<150\gev$.

Such low values of $\mueff$ are very favorable in point of view of finetunig, in particular if stops are also light. Indeed a good fraction of our points with degenerate $\hi,\hii$ and $R(\gam\gam)>1$ features light stops with $M_{\rm SUSY}=\sqrt{\mstopi\mstopii}\lesssim 1$~TeV. The stop mixing is typically large in these cases, $(A_t-\mueff\cotb)/M_{\rm SUSY}\approx 1.5\mbox{--}2$. Moreover, the few points which we found in the WMAP window always have $\mstopi<700\gev$. 

\begin{figure}[ht]\centering
\hspace*{-.25in}\includegraphics[width=0.54\textwidth]{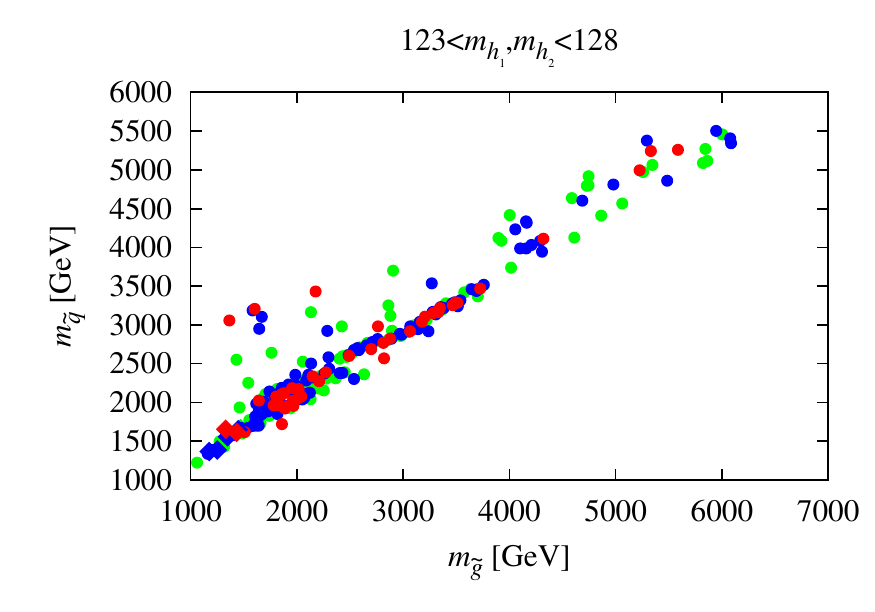}
\vspace*{-.1in}
\caption{Average light-flavor squark mass, $\msq$, versus gluino mass, $\mgl$, for the points plotted in the previous figures.}
\label{glsq}
\end{figure}

Implications of the enhanced $\gam\gam$ rate scenarios for other observables are also quite interesting. First, let us observe from Fig.~\ref{glsq} that these scenarios have squark and gluino masses that are above about $1.25\tev$ ranging up to as high as $6\tev$ (where our scanning more or less ended).  The WMAP-window points with large $R^h_{gg}(\gam\gam)$ are located at low masses of $\mgl\sim 1.3\tev$ and $\msq\sim 1.6\tev$.

\begin{figure}[ht]
\begin{center}
\hspace*{-.22in}\includegraphics[width=0.52\textwidth]{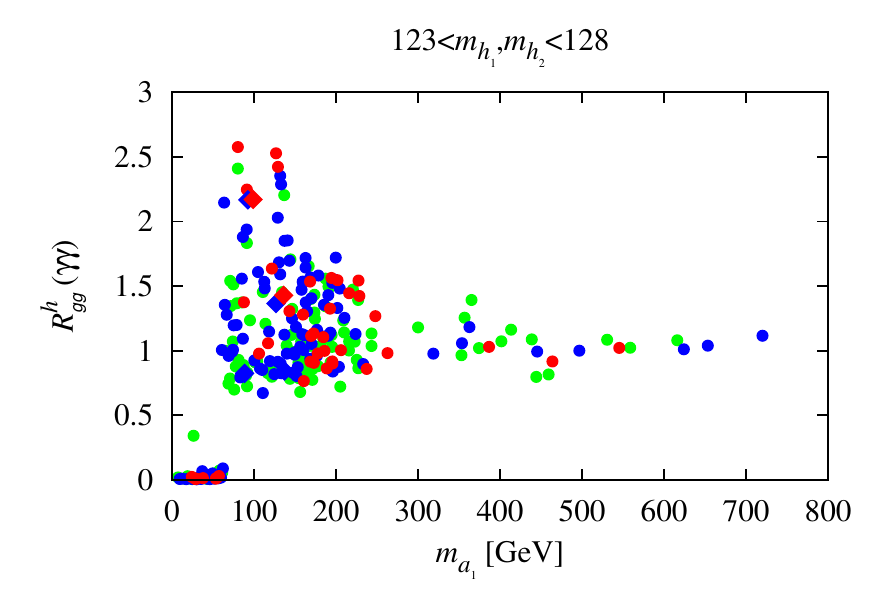}\hspace*{-.2in}\includegraphics[width=0.52\textwidth]{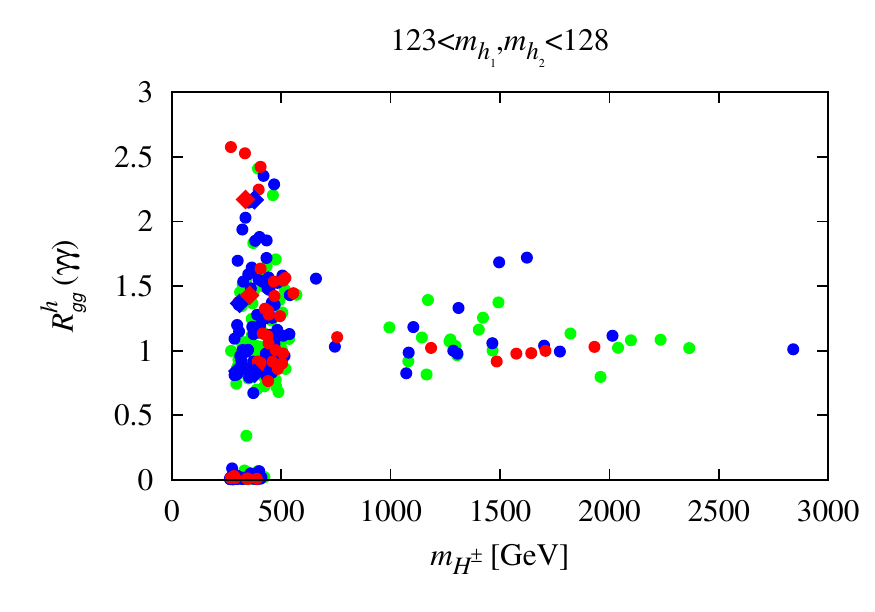}
\end{center}\vspace*{-.1in}
\caption{$R^h_{gg}(\gam\gam)$ versus the masses of $\mai$ and $\mhpm$ (note that $\mhpm\simeq\maii\simeq\mhiii$).}
\label{rvshmasses}
\end{figure}

The value of $R^h_{gg}(\gam\gam)$ as a function of the masses of the other Higgs bosons is illustrated in
Fig.~\ref{rvshmasses}.  We see that values above $1.7$ are associated with masses for the $\aii$, $\hiii$ and $\hpm$ of order $\lsim 500\gev$ and for the $\ai$ of order $\lsim 150\gev$.  (Note that $m_{\aii} \simeq m_{\hiii} \simeq m_{\hpm}$) While modest in size, detectability of these states at such masses requires further study. One interesting point is that $\mai\sim 125\gev$ is common for points with $R^h_{gg}(\gam\gam)>1$ points. We have checked that $R^{\ai}_{gg}(\gam\gam)$ is quite small for such points --- typically $\lsim 0.01$.

\begin{figure}[ht]
\begin{center}
\hspace*{-.2in}\includegraphics[width=0.52\textwidth]{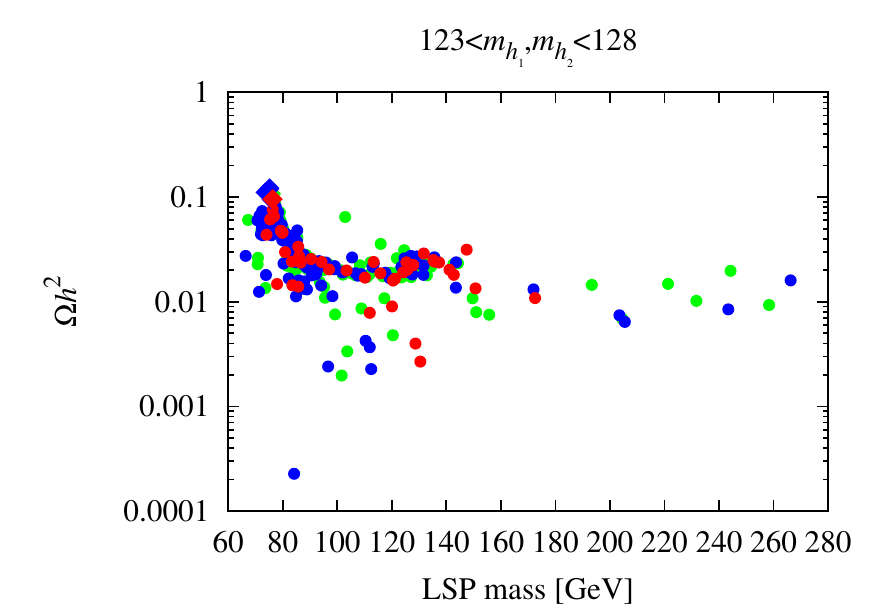}\hspace*{-.1in}\includegraphics[width=0.52\textwidth]{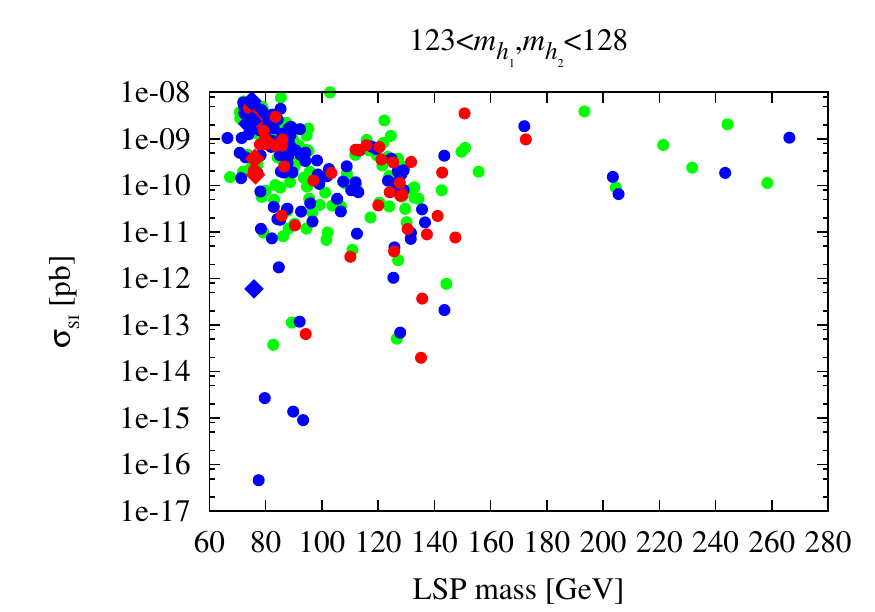}\\
\hspace*{-.2in}\includegraphics[width=0.52\textwidth]{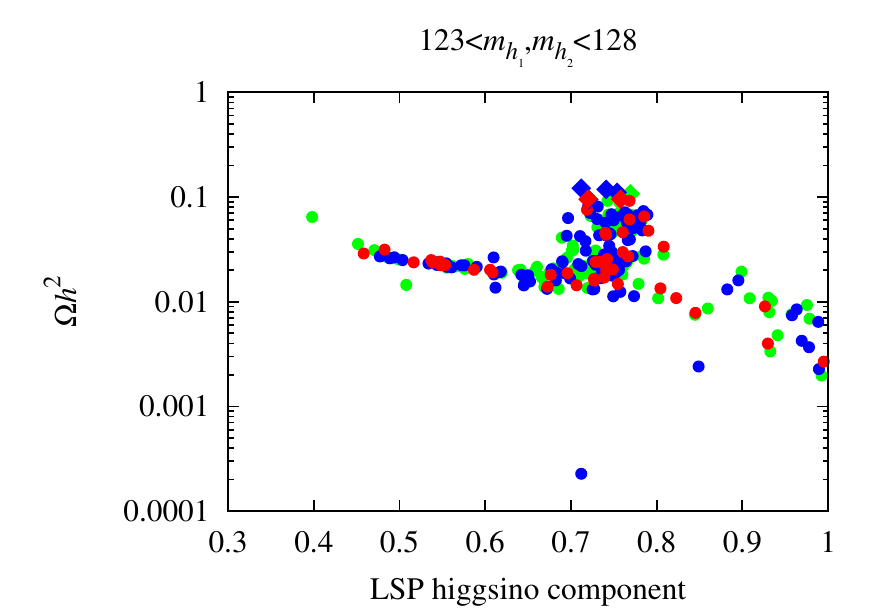}\hspace*{-.1in}\includegraphics[width=0.52\textwidth]{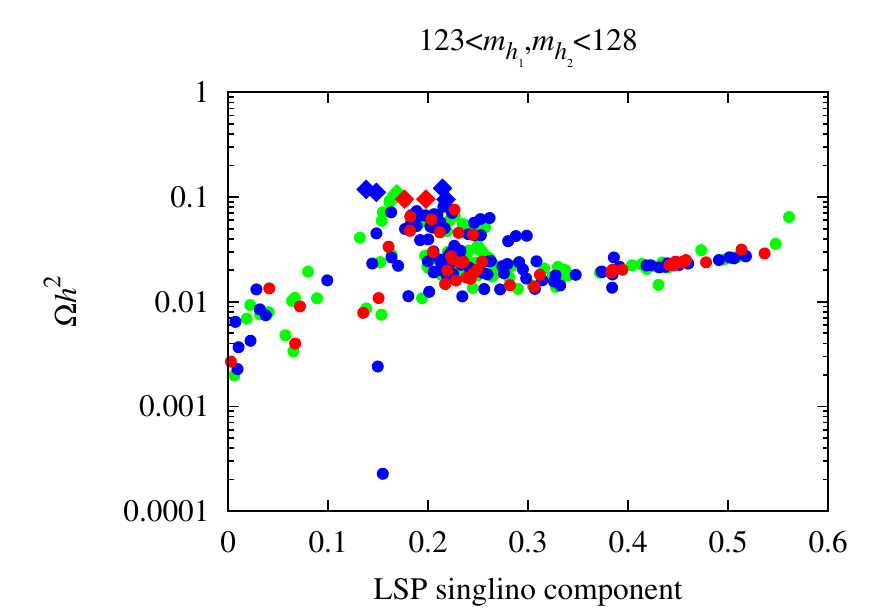}
\end{center}\vspace*{-.1in}
\caption{Top row: $\omghsq$ and spin-independent cross section on protons versus LSP mass for the points plotted in previous figures. Bottom row: $\omghsq$ versus LSP higgsino (left) and singlino (right) components.}
\label{lspfigs}
\end{figure}

Let us now focus on properties of the LSP.  In the plots of Fig.~\ref{lspfigs}, we display $\omghsq$ and the spin-independent cross section for LSP scattering on protons, $\sigsi$,  for the points plotted in previous figures.  We first note the rather limited range of LSP masses consistent with the WMAP window, roughly  $\mcnone\in[60,80]\gev$. The corresponding $\sigsi$ values show a broader range from $f\!ew\times 10^{-9}\pb$ to as low as $f\!ew\times 10^{-11}\pb$.  Owing to the small $\mueff$, the LSP is dominantly higgsino, which is also the reason for $\omghsq$ typically being too low. The points  with $\omghsq$ within the WMAP window are mixed higgsino--singlino, with a singling component  of the order of 20\%, see the bottom-row plots of Fig.~\ref{lspfigs}.

It is interesting to note a few points regarding the GUT-scale parameters associated with the points plotted in previous figures. For the WMAP-window diamond points, $m_0\in [0.9,1.3]\tev$, $\mhalf\in [500,700]\gev$, $A_0\in[ -1.8 , -1.0]\tev$, $\akap\in[-400,-250]\gev$, $\alam\in[ -600,-400]\gev$, $m_S(\text{GUT})\in [1.4,2.2]\tev$, $m_{H_u}(\text{GUT}) \in[ 2,2.2] \tev$ and $m_{H_d}(\text{GUT})\in [0.7,1.2]\tev$; and, as shown in earlier figures, these diamond points have $\lam\in[0.58,0.65]$, $\kappa\in[0.28,0.35]$, and $\tanb\in[2.5,3.5]$. Points with $R^h_{gg}(\gam\gam)>1.3$ have
$m_0\in [0.65,3]\tev$, $\mhalf\in [0.5,3]\tev$, $A_0\in[ -4.2 , -0.8]\tev$, $\akap\in[-500,+450]\gev$, $\alam\in[ -750,+550]\gev$, $m_S(\text{GUT})\in [1.2,4.2]\tev$, $m_{H_u}(\text{GUT}) \in[ 1.7,17] \tev$, $m_{H_d}(\text{GUT})\in [\sim 0,4.2]\tev$, $\lam\in[0.33,0.67]$, $\kappa\in[0.22,0.36]$, and $\tanb\in[2,14]$.

We have already noted that it is not possible to find scenarios of this degenerate/enhanced type while predicting a value of $\damu$ consistent with that needed to explain the current discrepancy.  In particular,  the very largest value of $\damu$ achieved is of order $1.8\times 10^{-10}$ and, further, the WMAP-window points with large $R_{gg}^h(\gam\gam,VV)$ have $\damu<6\times 10^{-11}$.


{\bf To summarize}, we have identified a set of interesting NMSSM scenarios in which the two lightest CP-even Higgs bosons are closely degenerate and lie in the $123\mbox{--}128\gev$ mass window.  Large rates (relative to $gg\to\hsm\to\gam\gam$ or $gg\to \hsm \to ZZ^*\to 4\ell$) for $gg\to h_{1,2} \to \gam\gam$ and $gg\to h_{1,2}\to ZZ^*\to 4\ell$ are possible, sometimes because one of the rates is large but also sometimes because the rates are comparable and their sum is large. This suggests that, especially if enhanced rates continue to be observed in these channels,  it will be important for the experimental community to be on the lookout for mass peaks in $m_{\gam\gam}$ and $m_{4\ell}$ that are broader than expected purely on the basis of the experimental mass resolution.  In addition, the apparent mass in the $\gam\gam$ final state might differ slightly from the apparent mass in the $4\ell$ final state. Significant statistics will be required to resolve such features.

\section*{Acknowledgements} 

This work originated from the workshop on ``Implications of a 125 GeV Higgs boson'' held at LPSC Grenoble from 30 Jan to 2 Feb 2012. We thank the other workshop participants, in particular U.~Ellwanger and G.~Belanger, for interesting discussions related to this study.  

This work has been supported in part by US DOE grant DE-FG03-91ER40674 and by  
IN2P3 under contract PICS FR--USA No.~5872.


\end{document}